\documentclass[aps,prl,twocolumn,superscriptaddress]{revtex4}

\usepackage{graphicx}
\begin{document}
\title{Valence evaluation of LiMnO$_2$ and related battery 
materials by x-ray absorption spectroscopy}

\author{H.~Wadati}
\email{wadati@phas.ubc.ca}
\homepage{http://www.geocities.jp/qxbqd097/index2.htm}
\affiliation{Department of Physics and Astronomy, 
University of British Columbia, 
Vancouver, British Columbia V6T 1Z1, Canada}

\author{D. G. Hawthorn}
\affiliation{Department of Physics and Astronomy, 
University of Waterloo, 
Waterloo, Ontario N2L 3G1, Canada}

\author{T.~Z.~Regier}
\affiliation{Canadian Light Source, 
University of Saskatchewan, 
Saskatoon, Saskatchewan S7N 0X4, Canada}

\author{G.~Chen}
\affiliation{Department of Materials Science, 
College of Materials Science and Engineering, 
Jilin University, Changchun 130012, 
People's Republic of China}

\author{T.~Hitosugi}
\affiliation{WPI Advanced Institute for Materials Research (WPI-AIMR), 
Tohoku University, Sendai 980-8577, Japan}

\author{T.~Mizokawa}
\affiliation{Department of Complexity Science and Engineering, 
University of Tokyo, Kashiwa, Chiba 277-8561, Japan}

\author{A.~Tanaka}
\affiliation{Department of Quantum Matters, ADSM, 
Hiroshima University, Hiroshima 739-8530, Japan}

\author{G.~A.~Sawatzky}
\affiliation{Department of Physics and Astronomy, 
University of British Columbia, 
Vancouver, British Columbia V6T 1Z1, Canada}


\date{\today}
\begin{abstract}
We present an x-ray absorption study of the oxidation states 
of transition-metal-ions of LiMnO$_2$ and its related materials, 
widely used as cathodes in Li-ion batteries. 
The comparison between the obtained spectrum and 
the configuration-interaction cluster-model 
calculations showed that 
the Mn$^{3+}$ in LiMnO$_2$ is a mixture 
of the high-spin and low-spin states. 
We found that Li deficiencies occur in the case of Cr substitution, 
whereas there are no Li deficiencies in the case of Ni substitution. 
We conclude that the substitution of charge-transfer-type Ni or Cu 
is effective for LiMnO$_2$ battery materials. 
\end{abstract}
\maketitle
Many current electrodes of rechargeable Li batteries have 
layered structures. Among them, a layered transition-metal 
oxide Li$_x$CoO$_2$, in which the concurrent insertion 
of Li in a crystal structure and the reduction 
of the Co ions is achieved \cite{Mizushima}, 
has become widely used as a cathode electrode 
in the batteries in most portable electronics. 
Since then, much effort has been made to enhance 
charge/discharge rates of the cathode material 
and/or to replace Li$_x$CoO$_2$ by less expensive 
Li$_x$MnO$_2$ or related materials. 
In addition, in the recent ab-initio computational modeling study, 
LiNi$_{0.5}$Mn$_{0.5}$O$_2$ is proposed as a next generation 
cathode material with high power and high capacity 
applications \cite{KangSci}. Although the valence 
states and local electronic configurations of the transition-metal ions 
are related to the charge/discharge rates, no systematic 
spectroscopic study has been performed to identify 
their local electronic configurations experimentally. 
Here, we report a systematic x-ray absorption spectroscopy (XAS) 
study of LiMnO$_2$, LiNi$_{0.5}$Mn$_{0.5}$O$_2$, and related compounds 
which are thought as future cathode materials. 
The present experiment and theoretical analysis reveal that
LiMnO$_2$ has the Mn$^{3+}$ site with the unexpected degeneracy 
between the low-spin (LS) and high-spin (HS) states 
and that LiNi$_{0.5}$Mn$_{0.5}$O$_2$ consists of HS 
Ni$^{2+}$ and Mn$^{4+}$. 
In the case of LiMnO$_2$, small perturbation 
such as Cr doping may stabilize the LS Mn$^{3+}$ state. 
These results show the potential of LiMnO$_2$ and 
LiNi$_{0.5}$Mn$_{0.5}$O$_2$ for use 
as cathode materials and provide 
fundamental information that can be used for the 
development of new Li-ion batteries 
for high power applications. 

\begin{figure}
\begin{center}
\includegraphics[width=8cm]{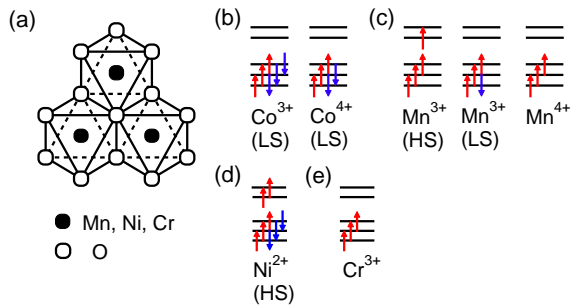}
\caption{(Color online) 
Crystal structure of cathode materials (a) and 
electronic configurations. 
(b) Co$^{3+}$ (LS) and Co$^{4+}$ (LS). 
(c) Mn$^{3+}$ (HS), Mn$^{3+}$ (LS) and Mn$^{4+}$. 
(d) Ni$^{2+}$ (HS). 
(e) Cr$^{3+}$. }
\label{fig1}
\end{center}
\end{figure}

LiCoO$_2$, LiMnO$_2$, and LiNi$_{0.5}$Mn$_{0.5}$O$_2$ 
have the layered structure as schematically shown 
in Fig.~\ref{fig1} (a). In this structure, 
the O$^{2-}$ ions provide 
a framework with face-centered cubic structure 
and Li$^+$ and transition-metal ions occupy 
all the octahedral interstices, leading to 
alternate two-dimensional triangular lattice layers. 
Under the octahedral crystal field, the five-fold transition-metal 
$3d$ level is split into two-fold $e_g$ and three-fold $t_{2g}$  
levels. In these layered cathode materials, the Li$^+$ 
ions diffuse between the octahedral sites 
through the tetrahedral interstices. In LiCoO$_2$, the pathways of 
Li$^+$ ions are in contact with the LS Co$^{3+}$ 
in which the $t_{2g}$ orbitals are fully occupied ($t_{2g}^6$). 
After the charging process, the LS Co$^{3+}$ ($t_{2g}^6$) 
is partly changed into the LS Co$^{4+}$ ($t_{2g}^5$), 
in which the structural change by the oxidation is minimal. 
The electronic configurations of these two Co ions are 
shown in Fig.~\ref{fig1} (b). 
If the Mn$^{3+}$ ion in LiMnO$_2$ and the Mn$^{4+}$ ion 
after the charging process take the electronic configurations 
of LS $t_{2g}^4$ and $t_{2g}^3$, respectively, 
as recently indicated by ab-initio calculation \cite{limno1}, 
the structural change by the oxidation is also expected to be minimal. 
If the Mn$^{3+}$ ion in LiMnO$_2$ takes the HS configuration 
with one $e_g$ electron, the Mn octahedral site should be 
deformed by Jahn-Teller effect to lift the two-fold degeneracy 
of the $e_g$ levels. In such case, local structural change 
during the charging process is substantial, leading to poor 
charge/discharge rate as well as poor charge/discharge cycle lifetime. 
The electronic configurations of these three Mn ions are 
shown in Fig.~\ref{fig1} (c). 
In LiNi$_{0.5}$Mn$_{0.5}$O$_2$, half of the Li$^+$ activated sites 
are in contact with Ni$^{2+}$ and higher Li diffusivity is expected. 
If the Ni$^{2+}$ ion in LiNi$_{0.5}$Mn$_{0.5}$O$_2$ and 
the Ni$^{4+}$ ion after the charging process take 
the HS configuration $t_{2g}^6e_g^2$ and the LS 
configuration $t_{2g}^6$, respectively, both of 
these ions are Jahn-Teller inactive. 

As discussed in the previous paragraph, the local electronic 
configuration of the transition-metal ions is highly important 
to determine the electrochemical properties 
of the cathode materials. In this report, we present 
a systematic XAS study of the cathode materials 
to extract fundamental information 
on their electronic configurations. 
We determined the valence of transition-metal ions 
by XAS and discussed what type of substitution is 
most suitable for LiMnO$_2$ batteries. 

The powder samples of LiMnO$_2$, 
LiMn$_{0.5}$Ni$_{0.5}$O$_{2}$, and 
LiMn$_{0.65}$Cr$_{0.35}$O$_2$ were 
synthesized by the procedure described in 
Refs.~\cite{limno1, limno2}. 
The powder of 
LiMnO$_2$ and LiMn$_{0.5}$Ni$_{0.5}$O$_{2}$ 
have a monoclinic structure, 
while LiMn$_{0.65}$Cr$_{0.35}$O$_2$ 
has a rhombohedral ($\alpha$-NaFeO$_2$-type) one. 
XAS experiments were performed 
at 11ID-1 (SGM) of the Canadian Light Source. 
The spectra were measured in the total-electron-yield 
(TEY) mode. The obtained TEY spectra were similar to 
the partial-fluorescence-yield spectra. 
The resolution power ($E/\Delta E$) was set to 5000. 
All the spectra were measured at room temperature. 

\begin{figure}
\begin{center}
\includegraphics[width=7cm]{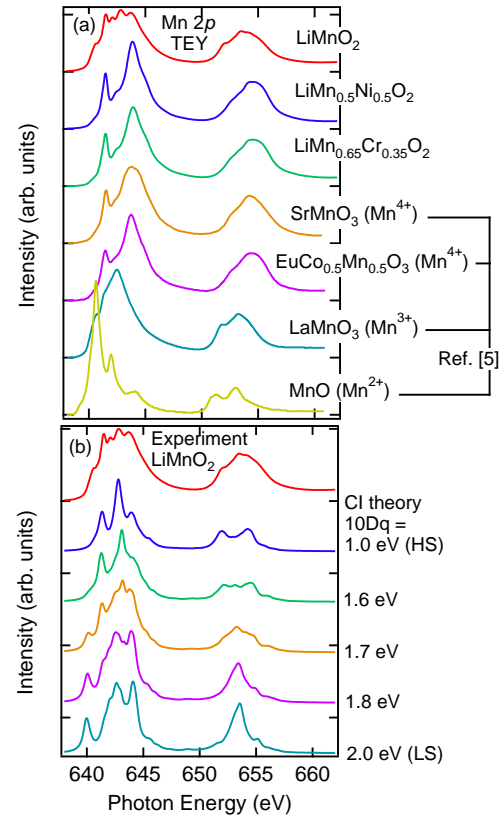}
\caption{(Color online) 
(a) Mn $2p$ XAS spectra and their comparison 
with the reference data of Mn$^{2+}$ (MnO), 
Mn$^{3+}$ (LaMnO$_3$), and Mn$^{4+}$ 
(EuCo$_{0.5}$Mn$_{0.5}$O$_3$ and SrMnO$_3$) 
from Ref.~\cite{MnCo}. 
(b) Comparison with the CI theory.}
\label{fig2}
\end{center}
\end{figure}

Figure \ref{fig2} (a) shows the Mn $2p$ XAS spectra and 
their comparison with the reference data of 
Mn$^{2+}$ (MnO), Mn$^{3+}$ (LaMnO$_3$), and Mn$^{4+}$ 
(EuCo$_{0.5}$Mn$_{0.5}$O$_3$ and SrMnO$_3$) 
from Ref.~\cite{MnCo}. 
The spectra of LiMn$_{0.5}$Ni$_{0.5}$O$_{2}$ and 
LiMn$_{0.65}$Cr$_{0.35}$O$_2$ 
are very similar to those of 
the Mn$^{4+}$ references, indicating that the valence of 
Mn is $4+$ in these materials. The spectrum of 
LiMnO$_2$ is different from the other three 
results and also from the Mn$^{2+}$ and Mn$^{4+}$ 
references. It is similar to the Mn$^{3+}$ reference of 
LaMnO$_3$, but the lineshapes are slightly different. 
We considered that this difference may come from 
the spin states, that is the difference between 
HS and LS states, and performed 
configuration-interaction (CI) cluster-model 
calculations \cite{atanaka2}. Figure \ref{fig2} 
(b) shows the comparison with the CI theory. 
The value of the 
crystal field splitting between $e_g$ and 
$t_{2g}$ states, $10Dq$, was changed 
from $1.0$ eV to $2.0$ eV. The positive value of $10Dq$ means 
that $e_g$ states have a higher energy than 
$t_{2g}$, which is the case in the MnO$_6$ 
octahedral coordination as shown in Fig.~\ref{fig1} (a) \cite{foot}. 
The increase of $10Dq$ from $1.0$ eV to 
$2.0$ eV corresponds to the transition from HS to 
LS states. (The electron configurations are 
shown in Fig.~\ref{fig1} (b)). 
The spectrum of LaMnO$_3$ in Fig.~\ref{fig2} (a) is in 
good agreement with the results of 
CI theory with $10Dq=1.0$ or 1.6 eV, which shows that Mn$^{3+}$ in 
LaMnO$_3$ is in the HS state, consistent with 
the reported result \cite{Abbate}. 
The spectrum of LiMnO$_2$ in Fig.~\ref{fig2} (a) 
is in good agreement with the CI theory of 
$10Dq=1.7$ eV. This result demonstrates that 
the Mn$^{3+}$ in LiMnO$_2$ is a mixture of 
the HS and LS states, consistent with 
the result from a similar analysis 
by de Groot \cite{grootrev}. If the $10Dq$ value 
of LiMnO$_2$ is slightly increased by chemical substitution, 
substrate strain etc., then the Mn$^{3+}$ LS state 
can be realized in LiMnO$_2$, as predicted by 
first-principle calculations \cite{limno1}. 

\begin{figure}
\begin{center}
\includegraphics[width=9cm]{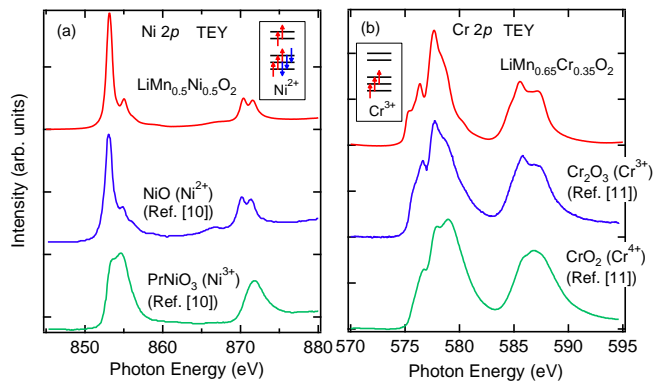}
\caption{(Color online) 
(a) Ni $2p$ XAS spectrum of LiMn$_{0.5}$Ni$_{0.5}$O$_{2}$ 
and its comparison with the reference data of 
Ni$^{2+}$ (NiO) and Ni$^{3+}$ (NdNiO$_3$ and PrNiO$_3$) 
from Ref.~\cite{Medarde}. 
The inset shows the electron configuration of Ni$^{2+}$. 
(b) Cr $2p$ XAS spectrum of LiMn$_{0.65}$Cr$_{0.35}$O$_2$ 
and its comparison with the reference data of 
Cr$^{3+}$ (Cr$_2$O$_3$) and Cr$^{4+}$ (CrO$_2$) 
from Ref.~\cite{Dedkov}. 
The inset shows the electron configuration of Cr$^{3+}$.}
\label{fig3}
\end{center}
\end{figure}

We also determined the valence of Ni and Cr by XAS spectra. 
Figure \ref{fig3} shows the Ni $2p$ XAS spectrum 
of LiMn$_{0.5}$Ni$_{0.5}$O$_{2}$ 
and its comparison with the reference data of 
Ni$^{2+}$ (NiO) and Ni$^{3+}$ (NdNiO$_3$ and PrNiO$_3$) 
from Ref.~\cite{Medarde}. The spectrum of 
LiMn$_{0.5}$Ni$_{0.5}$O$_{2}$ is similar to that of 
NiO rather than NdNiO$_3$ and PrNiO$_3$, indicating that 
the valence of Ni is $2+$ in LiMn$_{0.5}$Ni$_{0.5}$O$_{2}$. 
Similarly, it is shown from the Cr $2p$ XAS 
spectrum of LiMn$_{0.65}$Cr$_{0.35}$O$_2$ Fig.~\ref{fig4} 
and its comparison with the reference data of 
Cr$^{3+}$ (Cr$_2$O$_3$) and Cr$^{4+}$ (CrO$_2$) 
from Ref.~\cite{Dedkov} that our result is similar to 
that of Cr$_2$O$_3$ and the valence of Cr is 
$3+$ in LiMn$_{0.65}$Cr$_{0.35}$O$_2$. 

From the above results, we determined that the 
electronic configurations in powder samples are  
LiMn$^{3+}$O$_2$, 
LiMn$^{4+}_{0.5}$Ni$^{2+}_{0.5}$O$_2$, 
and Li$_x$Mn$^{4+}_{0.65}$Cr$^{3+}_{0.35}$O$_2$. 
In the Cr-substituted sample, we have to take into 
account the Li deficiencies to keep the material 
charge-neutral, and the value of Li concentration 
$x$ is calculated to be $0.35$. Such a large 
Li deficiency is also concluded from 
the results of 
photoemission spectroscopy \cite{Takaiwa}. 

When we substitute Cr for Mn in LiMnO$_2$, 
the Cr$^{3+}$ valence and the change from 
Mn$^{3+}$ to Mn$^{4+}$ introduce Li 
deficiencies and make this material unsuitable 
for battery cathodes. 
In the case of Ni substitution, 
the Ni$^{2+}$ valence and the change from 
Mn$^{3+}$ to Mn$^{4+}$ do not create 
Li deficiencies and the material is suitable 
for cathodes. When we classify transition-metal 
compounds in the scheme of Zaanen, Sawatzky and Allen 
\cite{ZSA, earlyCI}, that is as a function of 
Coulomb repulsion $U$ and charge-transfer energy 
from O $2p$ states to transition-metal $2p$ states 
$\Delta$, charge-transfer-type ($\Delta < U$) 
such as Ni and Cu is more suitable for LiMnO$_2$ 
battery cathodes than Mott-Hubbard type ($U< \Delta$) 
such as Cr because Ni and Cu becomes $2+$ and 
does not cause Li-deficiencies in the material. 

We performed an x-ray absorption study of 
LiMnO$_2$ and its related materials and 
determined the oxidation states 
of transition-metal ions. 
The XAS result of LiMnO$_2$ shows that the Mn$^{3+}$ 
in LiMnO$_2$ is a mixture of the HS and LS states, 
suggesting that small perturbation such as Cr doping 
in powder samples or substrate strain of thin films 
may stabilize the LS Mn$^{3+}$ state in LiMnO$_2$. 
In Li$_x$Mn$_{0.65}$Cr$_{0.35}$O$_2$, 
although the substituted Cr becomes $3+$ as expected, 
the Cr doping unfortunately causes Li deficiencies 
and introduces Mn$^{4+}$. 
In contrast, LiMn$_{0.5}$Ni$_{0.5}$O$_2$ 
has substituted Ni of $2+$ and does not include Li 
deficiencies. 
Thus we conclude that the substitution of charge-transfer-type 
Ni or Cu is effective for LiMnO$_2$ battery materials. 

This research was made possible with 
financial support from the Canadian funding organizations 
NSERC, CFI, and CIFAR.


\begin{thebibliography}{10}
\bibitem{foot}
Based on the crystal structure of these materials, the Mn ion is in a trigonal symmetry, but since the trigonal distortion is small, 
we consider it a reasonable approximation to calculate using an 
octahedral symmetry. 
The parameters in the calculation are the charge-transfer 
energy from the O $2p$ orbitals to the empty Mn $3d$ orbitals 
$\Delta = 4.5$ eV, the $3d-3d$ on-site Coulomb interaction 
energy denoted $U=4.0$ eV, and the hybridization strength 
between the Mn $3d$ and O $2p$ orbitals $V(e_g)=3.0$ eV.
\end{thebibliography}
\end{document}